\documentclass{mem}
\usepackage{natbib}\usepackage{txfonts}\usepackage{balance}
\usepackage{graphicx}
\usepackage[a4paper,breaklinks,dvipdfm]{hyperref}
\idline{75}{282}
\begin{document}
\def\teff{$T\rm_{eff }$}
\def\kms{$\mathrm {km s}^{-1}$}

\title{
On the circumstellar medium of massive stars and how it may appear in GRB observations
}

   \subtitle{}

\author{
A.J.\,van  Marle\inst{1}
\and R.\,Keppens\inst{1}
\and S.-C.\,Yoon\inst{2}
\and N.\,Langer\inst{2}
          }

  \offprints{A.J. van Marle}

\institute{
Centre for Plasma Astrophysics, 
K.U. Leuven, Celestijnenlaan 200B,
3001 Heverlee, Belgium,
\email{AllardJan.vanMarle@wis.kuleuven.be}
\and
Argelander Institut f{\"u}r Astronomie, University of Bonn, Auf
dem H{\"u}gel 71, D-53121, Bonn, Germany
}

\authorrunning{van Marle et al.}

\titlerunning{CSM of GRB progenitors}

\abstract{Massive stars lose a large fraction of their original mass over the course of their evolution. 
These stellar winds shape the surrounding medium according to parameters that are the result of the characteristics of the stars, varying over
time as the stars evolve, leading to both permanent and temporary features that can be used to constrain the evolution of the progenitor star.

Because long Gamma-Ray Bursts (GRBs) are thought to originate from massive stars, the characteristics of the circumstellar medium (CSM) should be observable in the signal of GRBs.  
This can occur directly, as the characteristics of the GRB-jet are changed by the medium it collides with, and indirectly because the GRB can only be observed through the extended circumstellar bubble that surrounds each massive star. 

We use computer simulations to describe the circumstellar features that can be found in the vicinity of massive stars and discuss if, and how, they may appear in GRB observations. 
Specifically, we make hydrodynamical models of the circumstellar environment of a rapidly rotating, chemically near-homogeneous star, which is
represents a GRB progenitor candidate model. 

The simulations show that the star creates a large scale bubble of shocked wind material, which sweeps up the interstellar medium in an expanding shell. 
Within this bubble, temporary circumstellar shells, clumps and voids are created as a result of changes in the stellar wind.
Most of these temporary features have disappeared by the time the star reaches the end of its life, leaving a highly turbulent circumstellar bubble behind. 
Placing the same star in a high density environment simplifies the evolution of the CSM as the more confined bubble prohibits the formation of some of the temporary structures.

}
\maketitle{}

\section{Introduction}
As stars evolve, they lose mass in the form of stellar wind. In the case of massive stars ($\gtrsim12\,$M$_\odot$) these winds can be very strong due to their high luminosity, which drives the wind. 
As a stellar wind expands into the surrounding interstellar medium (ISM) it creates a large ($\gtrsim10\,$pc) bubble, with a complicated internal morphology reflecting the characteristics of the stellar wind, which in turn reflect the evolution of the progenitor star. 
The morphology of the circumstellar bubble can be observed directly (e.g. in the form of circumstellar nebula), but can also show up in observations of long  Gamma-Ray Bursts (GRBs), which can occur once the central star reaches the end of its life. 
This happens in two ways: 
\begin{enumerate}
\item In a GRB afterglow circumstellar gas may be observed as blue-shifted absorption features in the spectrum of the GRB afterglow \citep[e.g.][]{Prochaskaetal:2008}.
\item The circumstellar gas may influence the evolution of the GRB itself. 
As the GRB jet expands, it will collide with circumstellar gas, which, depending on the density of the circumstellar medium (CSM) can cause a change in the expansion velocity and lead to a change in the observed signal as kinetic energy is transferred to heat and then radiated away 
\citep[e.g.][]{vanEertenetal:2010}.
\end{enumerate}
In this paper we describe some of the features found in the circumstellar medium of massive stars and discuss if, and how, they may show up in the observations of GRBs. 

\section{Morphology of circumstellar bubbles}
\subsection{General shape}
The general (semi-permanent) shape of wind-blown bubbles was described analytically by \citet{Weaveretal:1977}. 
Close to the star the wind expands in all directions, leading to a $1/r^2$ density profile. 
Eventually this wind passes through the wind termination shock and enters the ``shocked wind bubble'', which has a (nearly) constant density. 
Since the kinetic energy of the wind is turned into heat at the termination shock, the shocked wind has a very high ($\sim10^7\,$K) temperature, which gives it a high thermal pressure, causing it to expand outwards, sweeping up a shell of shocked interstellar gas. 

\subsection{Temporary features}
As the star evolves, changes in the massloss rate and wind velocity cause  shells to appear inside the bubble. 
These  temporary features, which can be observed as circumstellar nebulae, have only a relatively short ($\sim10^4-10^5$\,yr) live span, before they dissipate. 
Therefore, they can be used to determine the evolution of the progenitor star and, if observed, can be powerful indicators as to what type of star produces a certain type of supernova or GRB.

Due to their more complicated morphology the evolution of such shells can best be described through numerical simulations. 
Such simulations have been made for the transition from red supergiant (RSG) to Wolf-Rayet (WR) star \citep{GarciaSeguraetal:1996b,vanMarleetal:2005,Freyeretal:2006} and luminous blue variable (LBV) to WR star \citep{GarciaSeguraetal:1996b,Freyeretal:2003,vanMarleetal:2007}. 
Recent work by \citet{ToalaArthur:2011} described similar interactions and included additional physics in the form of radiative transfer and thermal conduction, while \citet{Dwarkadas:2007} showed the interaction between a supernova and the wind bubble in which it originates. 
The effect of an anisotropic stellar wind, resulting from stellar rotation was included by \citet{Chitaetal:2008} and \citet{vanMarleetal:2008}. 

\begin{figure}[]
\resizebox{\hsize}{!}{\includegraphics[clip=true]{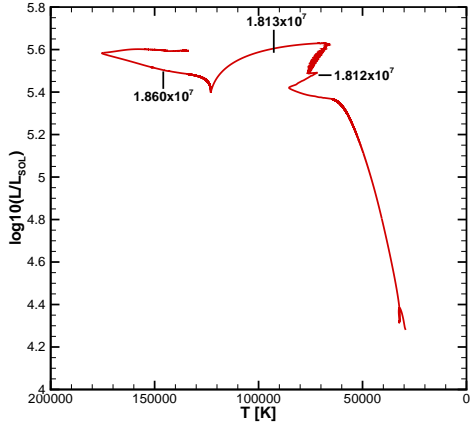}}
\caption{
\footnotesize
Hertzsprung-Russel diagram of a rapidly rotating 16\,M$_\odot$ star. The star has no giant stage and stays at the high temperature end of the diagram. 
The start and end of the first rapid rotation phase are marked, as is the start of the second rapid rotation phase. 
}
\label{fig:HR}
\end{figure}

\begin{figure}[]
\resizebox{\hsize}{!}{\includegraphics[clip=true]{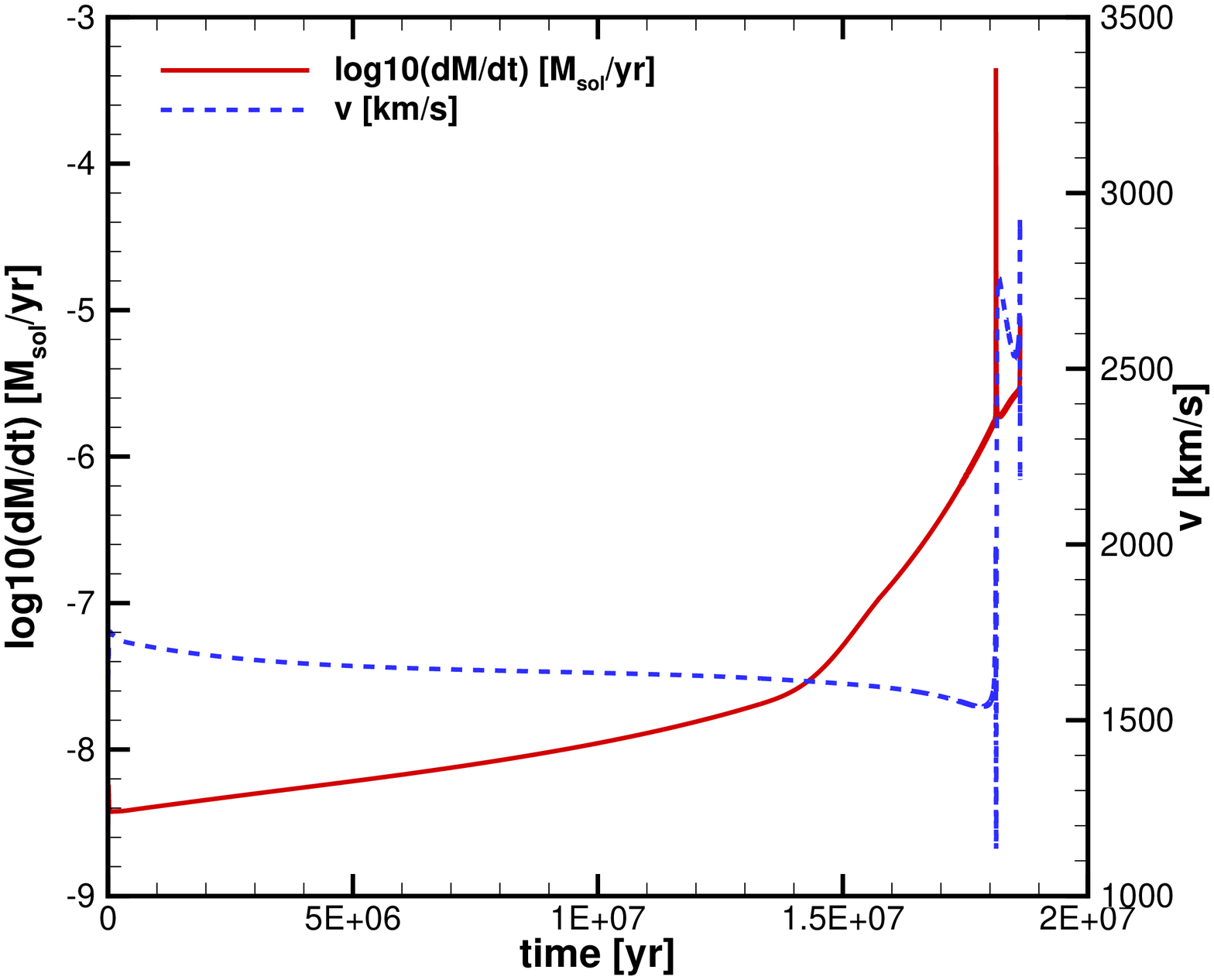}}
\caption{
\footnotesize
Mass loss rate and wind velocity (averaged over azimuthal angle) for the stellar evolution model from Fig.~\ref{fig:HR}. At the pole, massloss rate and wind velocity are higher than the average. At the pole, they are lower.
}
\label{fig:dm_v}
\end{figure}

\section{Numerical models}
\subsection{Stellar evolution models}
As input models for our numerical simulations we use one of the rapidly rotating, chemically near-homogeneous stellar evolution models from \citet{Yoonetal:2006}. These models follow special evolutionary tracks, which differ from the classical evolution due to their rapid rotation. 
Rotationally induced mixing allows the star to maintain hydrogen core burning until all hydrogen in the star is exhausted before it makes the transition to  helium burning. 
As a result the star has no giant phase, but makes a direct transition from main-sequence to WR phase. 
This transition (which occurs at the end of hydrogen burning) is marked by a short period during which the star rotates at critical velocity. 
Because such stars maintain a high rotation rate until the end of their existence, they are thought to be GRB progenitors. 
One such model was used in \citet{vanMarleetal:2008}. 
We now use the A16f0.5 model from the same paper to explore in more detail the interactions that take place in the circumstellar environment. 
This model represents a 16\,M$_\odot$ star with an initial metallicity of 0.002 and an initial rotation rate of $50\%$ of critical rotation. The evolutionary track of this star is shown in Fig.~\ref{fig:HR}. 
From this stellar evolution model we obtain massloss rate, wind velocity and rotation rate as a function of time (Fig.~\ref{fig:dm_v}), which are used as input for the hydrodynamical simulations. 
The episode of critical rotation, which marks the end of core hydrogen burning, is clearly visible as a massive increase in massloss rate at about $1.812\times10^7$\,yrs. 
Afterward, the massloss rate decreases again. 
A second outburst starts at the end of the stellar evolution ($\sim1.86\times10^7$\,yrs), but does not reach a similar height. 
During the outbursts, wind velocity is severely reduced. 

\subsection{Numerical hydrodynamics}
For our hydrodynamical simulations we use the {\tt MPI-AMRVAC} code \citep{Keppensetal:2011}. 
This code solves the conservation equations of mass, momentum and energy on an adaptive mesh grid and allows us to choose from a wide variety of solvers. 
We include optically thin radiative cooling, using the table from \citet{Schureetal:2009}, which was generated with the {\tt SPEX} package \citep{Kaastraetal:1996}. 

For the main sequence, during which no major changes in stellar wind parameters take place, we run the simulation in 1-D to safe computation time. 
We use a spherically symmetric grid with a maximum radius of $140\,$pc and a basic grid of 400 points. 
We allow 6 additional levels of refinement, so that our grid effectively has 25\,600 points. 
This grid is filled with a constant density of either 2 particles per cm$^{-3}$ (Simulation~A), representative of the general interstellar medium, or 2\,000 particles per cm$^{-3}$ (Simulation~B), representative of the dense molecular clouds in which massive stars form. 
At the inner radial boundary the grid is filled with matter according to the time-dependent stellar wind parameters obtained from the stellar evolution model. 

As the star approaches the critical rotation phase, it becomes necessary to model the circumstellar bubble in 2-D. 
Therefore, we map the 1-D result onto a 2-D grid. 
For this we choose a spherical grid in the r-$\theta$ plane, with a radius of $97.5\,$pc and a 90$^{\mathrm o}$ opening angle starting at the polar axis. 
The basic grid has 300$\times$200 points and is allowed two more levels of refinement, giving us an effective grid of 1200$\times$800 points. 
The azimuthal dependence of the massloss rate and wind velocity are calculated according to the gravity darkening model by \citet{Dwarkadasowocki:2002}, 
\begin{eqnarray}
 \dot{M}~&\propto&~1-\Omega^2\sin^2{\theta} \\
v_{\rm wind}~&\propto&~\sqrt{1-\Omega^2\sin^2{\theta}}
\end{eqnarray}
with $\dot{M}$ the massloss rate $v_{\rm wind}$ the radial wind velocity,$\Omega=v/v_{\rm crit}$ the rotation rate, relative to break-up speed and $\theta$ the co-azimuthal angle.
This model provides a more accurate description for radiatively driven winds than the older ``wind compressed disk'' model \citep{BjorkmanCassinelli:1993} used by \citet{Chitaetal:2008} and \citet{vanMarleetal:2008}. 
Because the gravity darkening model is only valid for the stellar wind and does not take into account the effect of matter being driven of the equator by centrifugal forces we truncate the rotation rate to $99{\%}$ of critical rotation. 

\begin{figure}[]
\resizebox{\hsize}{!}{\includegraphics[clip=true]{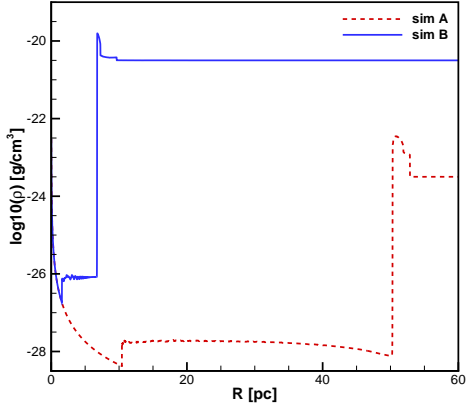}}
\caption{
\footnotesize
Density of the circumstellar medium at the end of the main sequence (after $1.81\times10^7$\,years) for both simulations A and B, just before the star reaches critical rotation. The solid line represents the simulation with high density ISM.
}
\label{fig:1D_dens}
\end{figure}

\begin{figure}[]
\resizebox{\hsize}{!}{\includegraphics[clip=true]{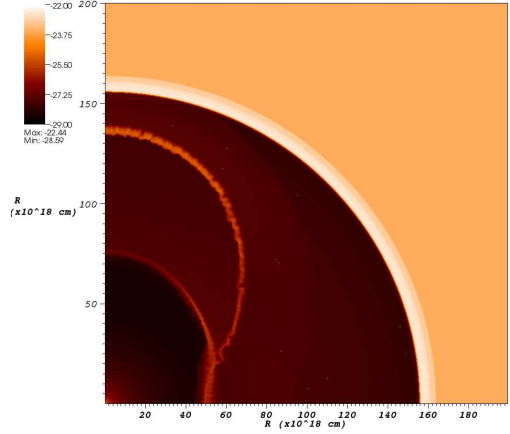}}
\caption{
\footnotesize
Circumstellar density for simulation~A $\sim60\,000$\,years after the result in Fig.~\ref{fig:1D_dens}, just after the critical rotation phase. During the critical rotation phase, which channels most of the wind toward the pole, the high massloss rate and low wind velocity cause the formation of a bipolar shell. Afterward, he wind velocity increases, leading to the formation of a third shell.
}
\label{fig:2D_low_0030}
\end{figure}

\begin{figure}[]
\resizebox{\hsize}{!}{\includegraphics[clip=true]{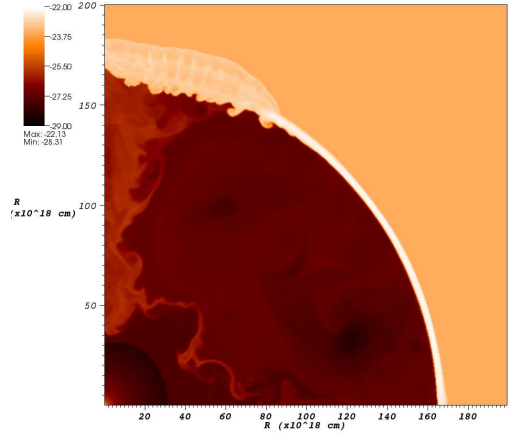}}
\caption{
\footnotesize
Similar to Fig.~\ref{fig:2D_low_0030} at  the end of the stellar evolution. The shells have disappeared, leaving a highly turbulent wind bubble.
}
\label{fig:2D_low_0269}
\end{figure}

\section{Results}
\subsection{1-D models}
At the end of the 1-D simulation ($1.81\times10^7$\,years after the start of the evolution), just before the star reaches critical rotation, the CSM morphology conforms to the analytical description from \citet{Weaveretal:1977}, as shown in Fig.~\ref{fig:1D_dens}. 
The effect of a high density ISM is clearly visible, as the bubble of simulation~B is far more compressed. 
This effect, which is predicted by the analytical approximation and shown numerically in \citet{vanMarleetal:2006}, is enhanced by radiative cooling, which is far more effective at high densities, causing energy to be lost from the system. 
the radiative cooling also causes the internal structure of the swept-up shell, which has the highest density at the contact discontinuity. 
As matter is swept up into the shell it starts to cool. 
The further it travels into the shell, the more it cools, causing to to reach its lowest temperature (and therefore highest density) at the contact discontinuity.

\subsection{2-D models}
As the star in Simulation~A reaches critical rotation, the wind velocity decreases and the massloss rate increases causing the formation of a bipolar shell, which moves outward into the hot bubble. 
The shape of the shell results from the azimuthal dependence of the wind, which directs most of the mass towards the pole \citep{Dwarkadasowocki:2002}.
Once the critical rotation phase ends, the wind velocity increases once again, causing the wind to sweep up its surroundings in a new shell, which will eventually overtake its bipolar predecessor (Fig.~\ref{fig:2D_low_0030}). 
This sequence of events in the CSM shows a close resemblance to the results from \citet{Chitaetal:2008} and \citet{vanMarleetal:2008}, despite the use of a completely different description for the wind of a rotating star. 
At the end of the stellar evolution (Fig.~\ref{fig:2D_low_0269}), both shells have disappeared after colliding with the outer shell and dissipating into the hot bubble. 
They leave behind a very turbulent region, especially over the pole of the star. 

\begin{figure}[]
\resizebox{\hsize}{!}{\includegraphics[clip=true]{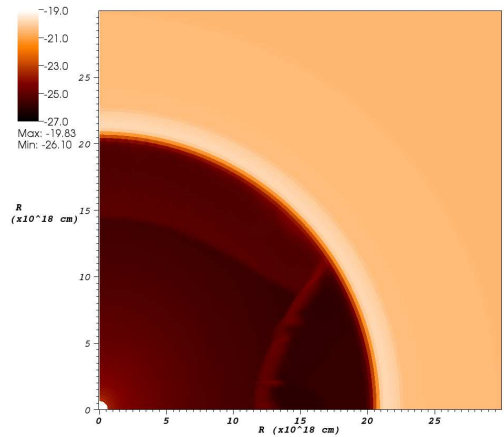}}
\caption{
\footnotesize
Circumstellar density for simulation~B~${\sim40\,000}$\,years after the result in Fig.~\ref{fig:1D_dens}. Because of the small size of the bubble, the bipolar shell has already collided with the outer shell in the polar region. At the equator the bipolar shell is still moving outward. The third shell never forms.
}
\label{fig:2D_high_0020}
\end{figure}

\begin{figure}[]
\resizebox{\hsize}{!}{\includegraphics[clip=true]{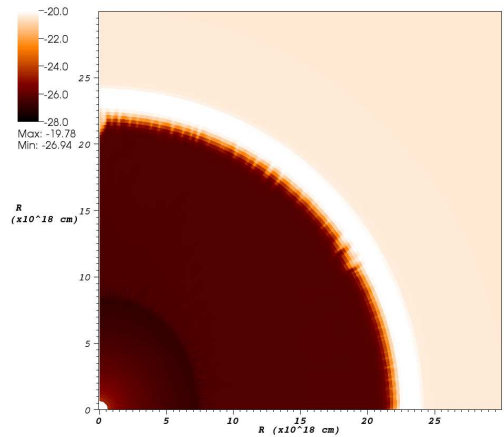}}
\caption{
\footnotesize
Similar to Fig.~\ref{fig:2D_low_0030} at the end of the stellar evolution. Unlike in simulation~A the hot bubble shows very little turbulence.
}
\label{fig:2D_high_0269}
\end{figure}

Figures~\ref{fig:2D_high_0020}-\ref{fig:2D_high_0269} show the CSM for simulation~B 40\,000\,years after the end of the 1-D simulation and at the end of evolution respectively. 
Because of the much smaller bubble, the bipolar shell that forms during critical rotation collides very early with the outer shell and disappears. 
The third shell never even has time to form. 
As a result, the CSM is much less turbulent at the end of the star's life.

Although the star spins up again at the end of its evolution, it does not again reach critical rotation and this episode has only minimal influence on the CSM, unlike the 20\,M$_\odot$ model described in \citet{vanMarleetal:2008}

% \begin{figure}[]
% \resizebox{\hsize}{!}{\includegraphics[clip=true]{A16f05_lowdens_0020.eps}}
% \caption{
% \footnotesize
% Density of the circumstellar medium for Simulation~A, $\sim40\,000$\,years after the result in Fig.~\ref{fig:1D_dens}  during the critical rotation phase , which channels most of the wind toward the pole. The high massloss rate and low wind velocity cause thr formation of a new, bipolar shell.
% }
% \label{fig:2D_high_0020}
% \end{figure}

% \begin{figure}[]
% \resizebox{\hsize}{!}{\includegraphics[clip=true]{A16f05_highdens_0018.eps}}
% \caption{
% \footnotesize
% Circumstellar density for simulation~B $\sim36\,000$\,years after the result in Fig.~\ref{fig:1D_dens}. Like in Simulation A, the critical rotation phase has formed a bipolar shell. 
% }
% \label{fig:2D_high_0018}
% \end{figure}

\section{Discussion and Conclusions}
The density of the ISM plays an important role in the evolution of the circumstellar bubble. 
The high density medium constrains the radius of the bubble, which in turn limits the formation of temporary features, such as circumstellar shells. 
Th effect of these limitations still shows at the end of the star's life, when the high density simulation shows much less turbulence than the more extended bubble in the low density medium. 

The temporary shells in the CSM, created by the changes in the stellar wind during the critical rotation phase, have disappeared by the time the star reaches the end of its evolution and the second phase was not powerful enough to generate a similar event. 
The critical rotation phase has left behind a great deal of turbulence in the shocked wind bubble. 
This results in density differences of more than order of magnitude that may influence the emission signal of the GRB afterglow as it sweeps up the high density regions. 

The absorption spectrum generated by this bubble should be unremarkable as the shells have disappeared, leaving only the free-streaming wind to generate a blue-shifted absorption feature \citep{vanMarleetal:2005,vanMarleetal:2007,vanMarleetal:2008} and this is so close to the star (and therefore the GRB), that only very high ionization states can survive 
\citep{Prochaskaetal:2007,Prochaskaetal:2008}. 
The outer shell might be visible as an independent absorption feature, but its velocity is low ($\sim25$\,km/s), making it nearly indistinguishable from the ISM.

\begin{acknowledgements}
A.J.v.M.\ acknowledges support from FWO, grant G.0277.08 and K.U.Leuven GOA/09/009. 
Simulations were done at the Flemish High Performance Computer Centre, VIC3 at K.U. Leuven. \\
\end{acknowledgements}

\bibliographystyle{aa}
\bibliography{vanmarle_biblio}
\IfFileExists{vanmarle_biblio.bbl}{}
 {\typeout{}
  \typeout{******************************************}
  \typeout{** Please run "bibtex \jobname" to obtain}
  \typeout{** the bibliography and then re-run LaTeX}
  \typeout{** twice to fix the references!}
  \typeout{******************************************}
  \typeout{}
 }

\end{document}